\documentclass{article}

\usepackage{PRIMEarxiv}

\usepackage[utf8]{inputenc} 
\usepackage[T1]{fontenc}    
\usepackage{hyperref}       
\usepackage{url}            
\usepackage{booktabs}       
\usepackage{amsfonts}       
\usepackage{nicefrac}       
\usepackage{microtype}      
\usepackage{lipsum}
\usepackage{fancyhdr}       
\usepackage{graphicx}       
\graphicspath{{media/}}     

\usepackage{natbib}
\usepackage{amsfonts} 
\usepackage{commath}
\usepackage{dsfont}
\usepackage{multirow}

\title{Joint Microseismic Event Detection and Location\\with a Detection Transformer
}

\author{
  Yuanyuan Yang, Claire Birnie, Tariq Alkhalifah\\
  Physical Science and Engineering Division\\
  King Abdullah University of Science and Technology\\
  \texttt{\{yuanyuan.yang, claire.birnie, tariq.alkhalifah\}@kaust.edu.sa} \\
}

\begin{document}
\maketitle

\begin{abstract}
Microseismic event detection and location are two primary components in microseismic monitoring, which offers us invaluable insights into the subsurface during reservoir stimulation and evolution. Conventional approaches for event detection and location often suffer from manual intervention and/or heavy computation, while current machine learning-assisted approaches typically address detection and location separately; such limitations hinder the potential for real-time microseismic monitoring. We propose an approach to unify event detection and source location into a single framework by adapting a Convolutional Neural Network backbone and an encoder-decoder Transformer with a set-based Hungarian loss, which is applied directly to recorded waveforms. The proposed network is trained on synthetic data simulating multiple microseismic events corresponding to random source locations in the area of suspected microseismic activities. A synthetic test on a 2D profile of the SEAM Time Lapse model illustrates the capability of the proposed method in detecting the events properly and locating them in the subsurface accurately; while, a field test using the Arkoma Basin data further proves its practicability, efficiency, and its potential in paving the way for real-time monitoring of microseismic events.
\end{abstract}


\section{Introduction}

From its inception in the 1970s and its later practical usage around 2000, microseismic monitoring has been helping us understand subsurface changes over time and processes that cause them \citep{warpinski2009microseismic}. While its most common use to date has been hydraulically-induced fracture mapping \citep{shapiro2006hydraulic}, it is also used for reservoir surveillance, $CO_2$ sequestration monitoring, seismic hazard assessment, and many other applications in the energy and mining industries \citep{maxwell2010petroleum, kaven2015surface, perol2018convolutional}. Among the information extracted from microseismic monitoring, the detection of microseismic events from the long seismic recordings and the estimation of their corresponding source locations are two primary components for observing, diagnosing, and acting upon the dynamic indications in the reservoir evolution and health. By tracking the fracturing properly, invaluable insights into the subsurface are gathered to optimize our strategies in, for example, enhanced geothermal systems (EGS) or carbon capture, utilization, and storage (CCUS) projects, and also to mitigate the environmental risks arising from induced seismicity \citep{young1992seismic, folesky2016rupture, li2022monitoring}. These important objectives call for exhaustive solutions to microseismic event detection and location.

Conventionally, microseismic event detection and location are typically addressed separately as two streams of tasks, and their corresponding methods are designed to be highly accurate or generic. However, for the majority of cases, such advantages come at a price, either requiring extensive human expertise or large computational overhead. For microseismic event detection, the short-term average to long-term average (STA/LTA) method \citep{allen1978automatic}, as one of the most widely used techniques for full-field detection, and most of its adaptations depend upon careful selection of parameters (such as window size and threshold) to reflect actual signal and noise conditions and complicated re-adjustment when the recording condition changes \citep{vaezi2015comparison, kumar2018development, zheng2018automatic}. Alternative detection methods like waveform template matching \citep{gibbons2006detection} have proven to be computationally intensive and time-consuming \citep{skoumal2016efficient}. For microseismic event location, traveltime-based methods \citep{waldhauser2000double, eisner2009uncertainties} can suffer from the heavy workload associated with manual traveltime picking, or can be prone to picking errors when done automatically, and thus require substantial time for post-processing quality control \citep{bose2009automatic, akram2016review}. Whereas migration-based \citep{mcmechan1982determination, larmat2006time, artman2010source} and inversion-based location methods \citep{sun2016full, wang2018microseismic, song2019passive} demand large computational memory and time for simulating the wavefields. The common requisite nowadays is to detect and locate the majority of microseismic events in a short time (i.e., real-time) in order to make instantly-informed decisions. Therefore, there is often a practicability gap towards the real-time field implementation of these methods.

Machine learning (ML) driven approaches can overcome the substantial limitations of manual intervention and/or high computational cost from conventional approaches and promise the possibility for real-time results in detecting and locating microseismic events, once the model is trained properly. They effectively front-load the computational burden to the offline time (i.e., the training stage), whereas during online operations (i.e., the inference stage), the trained models can fulfill the evaluations on the order of seconds \citep{grady2022model}. These approaches are therefore attractive for applications that require a large number of executions/iterations and demand real-time completion, since data generation and model training are covered at the offline stage. Much work has been in utilizing ML in microseismic monitoring, as summarized by \citet{anikiev2023machine}. ML-driven methods for microseismic event detection can be divided based on what is the input to ML algorithms, for example either a single trace or full arrays. Trace-based methods \citep[e.g.,][]{zheng2018automatic, chen2020automatic, wilkins2020identifying} take the input as a sequence. For example, \citet{birnie2022bidirectional} introduced a bidirectional recurrent neural network for accurate event detection via taking this task as binary classification for trace-wise recordings. The pitfall of such trace-based methods is that the arrays are not fully utilized, especially considering the poor signal-to-noise ratio (SNR) of microseismic data. On the contrary, array-based methods \citep[e.g.,][]{horne2019machine, birnie2021introduction, shaheen2021groningennet} treat the input as an image and thus can take advantage of array information like moveout patterns of energy across the sensors. \citet{stork2020application} provided one of the earliest successful examples, in which they trained the convolutional neural network (CNN) known as YOLOv3 to detect events based on the waveforms in array recordings. ML-driven methods for microseismic event location, similar to traditional location techniques, are mostly based on dense arrays of receivers. This is because microseismic data have less favorable SNR and are more sensitive to local heterogeneities compared to global seismology, which requires more information and features to locate the events. ML methods for event location utilize either the arrival times \citep{huang2018micro} or the full waveforms \citep{wamriew2020deep, wang2021data} to map the events to subsurface locations. There is also another interesting branch of research which uses the so-called physics-informed neural networks (PINNs) to locate microseismic events \citep{grubas2021localization, izzatullah2022laplace, huang2023microseismic}. These approaches, with very fast runtime during inference, are friendly to practical situations that require instant processing.

Some efforts have been dedicated to simultaneously detect and locate events within the seismogram segment \citep{drew2005automated, michaud2008continuous, cieplicki2012correlation, zhang2018automatic}. For example, \citet{wang2021direct} designed two CNNs: one of which aims to predict the number of events and the other is responsible for locating the present events. This method offers the opportunity of real-time computation when detecting and locating events. However, given that these two tasks are utilizing the same information (i.e., the waveforms), we believe it would be more beneficial to have a joint microseismic event detection and source location network that takes us from the recorded waveforms directly to the locations of the events responsible for these waveforms.

In this work, we unify microseismic event detection and location into a single framework by adapting the original DEtection TRansformer \citep{carion2020end}, which is based on a CNN backbone and a Transformer encoder-decoder general architecture with a set-based Hungarian loss. Using synthetic data from a 2D profile of the SEAM Time Lapse model and field data from the Arkoma Basin in North America, we demonstrate the robustness, efficiency, and transformative potential of the proposed approach for concurrent detection and location of microseismic events in real-time applications. The contributions of this work can be summarized as follows: (1) a pioneering introduction of an ML-based framework for joint microseismic event detection and location, trained offline on synthetic data; (2) leveraging the DEtection TRansformer network and a Hungarian loss to address the challenge of handling multiple events within a given recording segment; (3) conceptual proof and practical validation of the proposed method through successful applications on synthetic and field passive seismic recordings.

The subsequent sections of this paper are organized as follows: we start by describing essential elements of our approach, including the DEtection TRansformer model, its inputs and outputs, as well as the loss function computation; following this, we present the results on synthetic passive seismic data to assess the approach accuracy; a field data application highlights the advantages and limitations of the approach, which are further discussed in detail in the Discussion section; we conclude this study at the end.

\section{Methods}

Several ingredients are essential for joint microseismic event detection and location within a unified framework: (1) a network architecture that simultaneously predicts the number of events (a classification task) and the corresponding source locations of existing events (a regression task); (2) a series of data processing procedures to remove the need of source ignition time and to mitigate the discrepancies between synthetic (training) and field (inference) data; (3) a bipartite matching algorithm that ensures unique matching between ground-truth and predicted events for cases where there are multiple events within one input data segment; and (4) a set prediction loss function that utilizes the bipartite matching result and evaluates both detection and location predictions concurrently.

\subsection{The DEtection TRansformer model}

DEtection TRansformer (DETR) is a recently developed ML framework for object detection, which is a fundamental computer vision task aiming to detect object instances and locate them in images and/or videos. DETR has gained considerable attention due to its contribution towards eliminating the need for many hand-designed components while also demonstrating good performance \citep{dai2021up, liu2022petr, zhang2022accelerating}. Compared to classical CNN-based object detectors, DETR greatly simplifies the detection pipeline by exploiting the versatile and powerful relation-modeling capabilities of Transformers to replace hand-crafted rules including prior knowledge injection, like anchor generation, and post-processing, like non-maximum suppression \citep{zhu2020deformable, wang2021pnp, li2022dn}. The DETR architecture is powerful but conceptually simple; the following briefly describes the general framework, whilst we refer the readers to \citet{carion2020end} for a more in-depth discussion on the original DETR framework.

\begin{figure}
  \centering
  \includegraphics[width=1.0\textwidth]{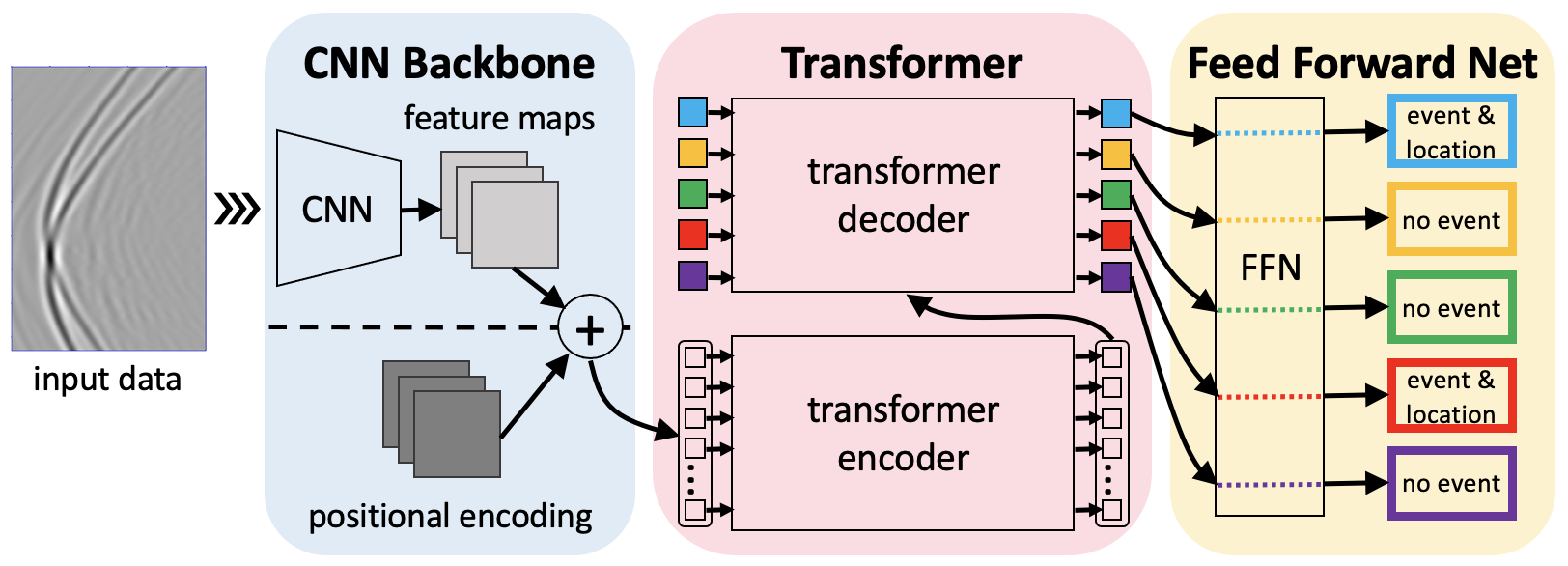}
  \caption{The architecture of DEtection TRansformer. The five output slots indicate that a maximum of five events can be detected and located per segment. In practice, this number should be set to a value significantly larger than the expected number of events within a single seismic segment to ensure that all potential events can be captured.}
  \label{fig:detr_architecture}
\end{figure}

As depicted in Fig. \ref{fig:detr_architecture}, DETR contains three main components: a conventional CNN backbone, an encoder-decoder Transformer, and a simple feed forward network (FFN). Fed with an input passive seismic data segment, the CNN generates a high-level activation map for extracting a compact feature representation. Considering the following Transformer architecture is permutation-invariant, DETR also supplements the feature maps with positional encodings to inject position information into the feature space.

Given the feature maps extracted from the CNN backbone, DETR leverages a standard Transformer encoder-decoder architecture to transform these feature maps into features of a set of prediction sequences using a multi-headed attention mechanism. The encoder expects sequences as inputs, therefore each two-dimensional feature map is flattened in the spatial dimension into a one-dimensional feature vector. For the decoder, the inputs include both output feature vectors from the encoder and $N$ input embeddings, where $N$ should be set to a value significantly larger than the typical number of expected events within one input passive seismic data segment. These input embeddings, shown as the colorful boxes fed to the Transformer decoder in Fig. \ref{fig:detr_architecture}, are learnable positional encodings that help the decoder extract event features. The decoder gathers feature information through the cross-attention modules to account for all potential events, while modeling the complicated interactions between features through the self-attention modules to prevent predicting duplicated events. In this way, the Transformer outputs $N$ prediction sequences in a single pass. In this work, we assume the maximum number of events, for test purposes, in the input windowed seismic segment is three, and thus, we set $N$ to five in its application.

\begin{figure}
  \centering
  \includegraphics[width=0.8\textwidth]{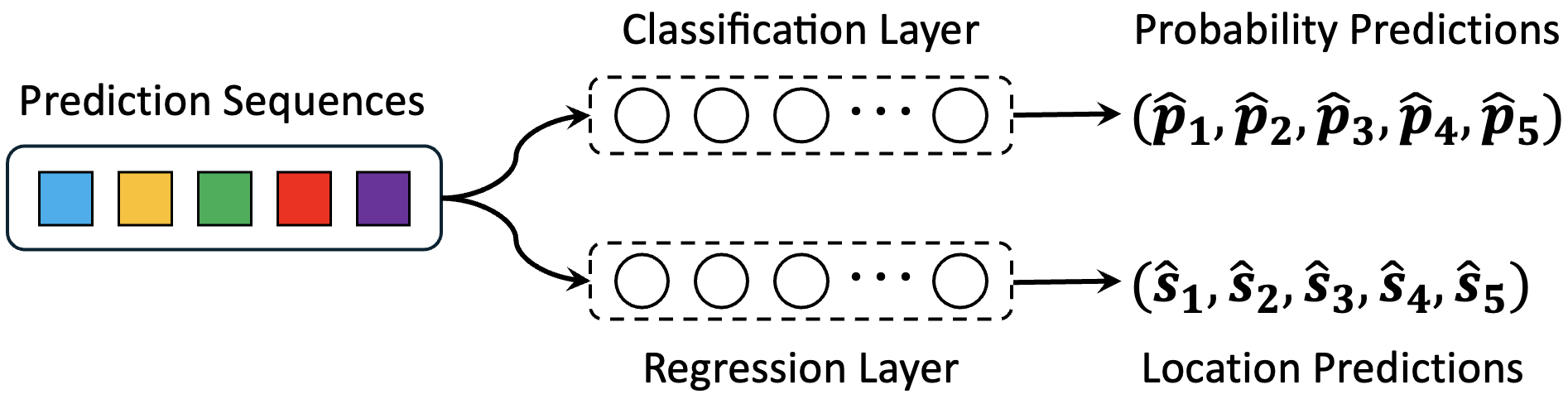}
  \caption{The diagram of Feed Forward Network branches. Fed with the prediction sequences from the Transformer, the classification branch produces the probability predictions $\hat{p}$ of event existence, while the regression branch produces the coordinate predictions $\hat{s}$ of corresponding source locations.}
  \label{fig:ffn_branches}
\end{figure}

The $N$ prediction sequences produced by the decoder are then passed to the FFN and independently decoded to make $N$ sets of final predictions. As shown in Fig. \ref{fig:ffn_branches}, there are two branches inside the FFN. One branch acts as a classification branch to produce the probabilities of event existence using a softmax activation function. Once the probability is larger than a determined threshold (e.g., 0.7 chosen in this work - as it was in the original DETR paper), an event is assumed to be detected in this slot. The other branch functions as a regression branch to predict the corresponding source coordinate locations within the area of interest in the subsurface model. To conclude, each set of final predictions incorporate a probability of an event existing within this prediction slot and a coordinate of the responsible source location.

\subsection{Network inputs: data generation and processing}

In this application, the input data to the DETR network are processed passive seismic recordings. To train the network, a forward simulation framework is required, where synthetic seismic data are first modeled with seismic events generated from random potential source locations using a velocity model that best represents the subsurface area of interest (e.g., a tomography inversion model). Due to the fact that microseismic events are generally weak in terms of energy and magnitude, for subsurface monitoring applications, only transmission P-wave components are typically used. Accordingly, we resort to the acoustic assumption and consider the locations of microseismic events as point sources. The acoustic wave equation offers a more efficient alternative, than a more realistic elastic wave equation, to generate the larger volume of passive seismic data containing microseismic events randomly ignited in the region of interest, which is needed for training the network. The suggested approach can be easily adaptable to an elastic or an anisotropic assumption of the Earth, which simply involves using the corresponding wave equation for modeling the training samples. However, it also requires knowledge of the corresponding medium parameters, like the shear wave velocity, or the elastic coefficients of the investigated area. In the proposed method, the network builds the projection from the patterns of microseismic energy across sensors to responsible source locations. This projection is based on the governing physics, which is implicitly embedded into the training data and corresponding labels. Therefore, the performance of this method, like any other locating method, is heavily dependent on the accuracy of the provided velocity model and physical assumptions used in the modeling.

Instead of directly using the lengthy seismic recordings as inputs to the network, the recordings are sliced into relatively short time segments and processed using the MLReal domain adaptation technique \citep{alkhalifah2022mlreal}. MLReal involves a series of efficient linear operations including cross-correlation, auto-correlation, and convolution between the synthetic and field data. These operations transform the features of the training dataset (i.e., the synthetic data) to incorporate those of the application dataset (i.e., the field data), and vice versa, bringing their data distributions much closer to each other. This can improve the generalization of the synthetically-trained ML models on field data. Without this domain adaptation, as observed in our initial experiments, the network trained on synthetic data struggles to generalize to field data due to their distributional discrepancy in data space. MLReal significantly mitigates this issue by reducing the distributional gap between the two datasets, which enhances network performance on field data application.

The new training data, $D_{train}(x, \tau)$, which is the synthetic passive seismic data transformed by the MLReal technique, can be defined as follows:
\begin{equation}
    D_{train}(x, \tau)= \left[ d_{syn}(x,t) \otimes d_{syn}(x=x_{ref},t) \right] \ast \left[ d_{fld}(x,t) \otimes d_{fld}(x,t) \right],
\label{eq:mlreal_transform_train}
\end{equation}
where $d_{syn}(x,t)$ is the synthetic seismic data segment, $d_{fld}(x,t)$ is a random sample of the field data segment, $x_{ref}$ is a fixed-location reference trace, $\otimes$ symbolizes the cross- or auto-correlation operation and $\ast$ represents the convolution operation. Specifically, the correlation operations convert the second axis from the time dimension $t$ to the time lag dimension $\tau$. In other words, Equation \ref{eq:mlreal_transform_train} represents the cross-correlation of the synthetic data segment with a fixed reference trace extracted from the segment itself and then the convolution of the resulting data with a random sample of the auto-correlated field data segment.

As for the new testing/application data, $D_{test}(x, \tau)$, the MLReal transformation is given by:
\begin{equation}
    D_{test}(x, \tau)= \left[ d_{fld}(x,t) \otimes d_{fld}(x=x_{ref},t) \right] \ast \left[ \frac{1}{m} \sum_{j=1}^m d_{syn}^j(x,t) \otimes d_{syn}^j(x,t) \right],
\label{eq:mlreal_transform_test}
\end{equation}
where $m$ is the total number of synthetic data segments in the training set, and $d_{syn}^j(x,t)$ denotes the $j^{th}$ synthetic segment in this training set. It includes the same operations as in Equation \ref{eq:mlreal_transform_train} with the role of synthetic and field data reversed. Furthermore, the mean of the auto-correlated synthetic data segments is utilized instead of a random sample.

\begin{figure}
  \centering
  \includegraphics[width=0.9\textwidth]{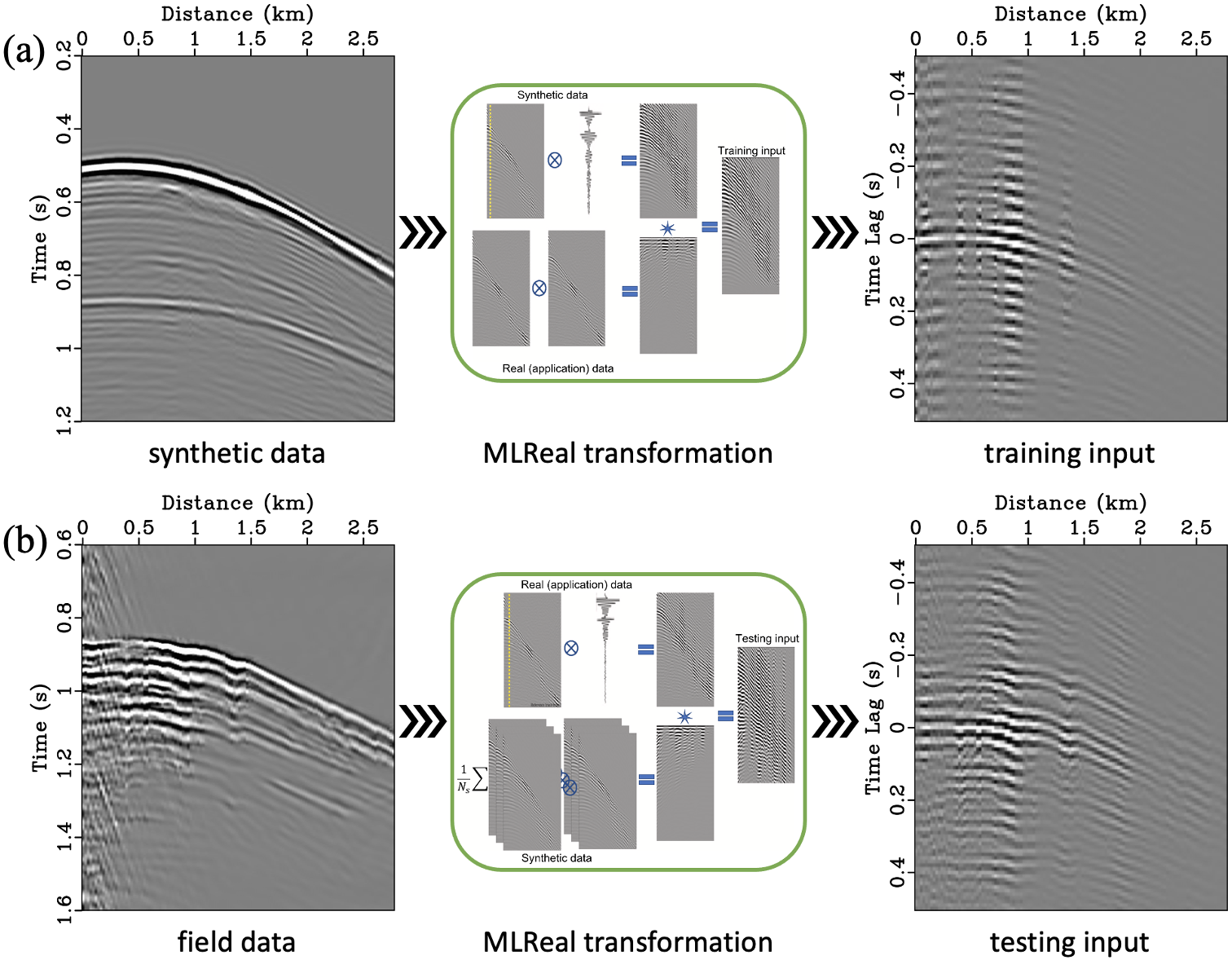}
  \caption{The MLReal transformation diagram for (a) training data and (b) application data, respectively.}
  \label{fig:mlreal_transform}
\end{figure}

The above MLReal domain adaptation technique helps to bridge the gap between training on synthetic data and applications on field data. Specifically, the MLReal transformation applied to the synthetic data transforms their feature space from the time domain to the time-lag domain through correlation operations, and incorporates realistic features from the field data (such as waveform signatures and noise conditions) into the synthetic training data via convolution. Whilst, the same MLReal transformation applied to the field data ensures that the field application data are projected to the same feature space (i.e., time-lag domain), and share a close distribution to the synthetic training data in this feature space (e.g., the similar frequency bands). This alignment of two data distributions allows the ML models to acquire strong generalization abilities to this specific label-less field data. Fig. \ref{fig:mlreal_transform} shows the MLReal transformation for both the training and application data. Note how similar the data features are after this domain adaptation. If the generated synthetic data are realistic enough, in which domain adaptation is not needed, the second part of Equations \ref{eq:mlreal_transform_train} and \ref{eq:mlreal_transform_test} (the auto-correlation and convolution operations) can be removed, which leaves us with only the cross-correlation part as the example provided in \citet{wang2021data}.

The correlation operation moves all the existing events within a recording segment to the zero-lag area, regardless of their ignition time. Therefore, in practice, segmentation of a fixed area around the zero lag and resampling can be applied to significantly reduce the input data size without losing useful event information (e.g., their curvature). In the interest of reducing the cost of generating synthetic data with varying numbers of events, we prepare the multiple-event synthetic training data by combining (stacking) modeled single-event data. This allows fast generation of a significant volume of training samples without the need for costly modeling of each multi-event scenario. This could be considered as a form of data augmentation.

\subsection{Network outputs: bipartite matching}

For each input data, the DETR network infers a set comprising $N$ elements, where $N$ is usually much larger than the expected maximum number of events in one input windowed seismic segment. Each element is a pair of predictions, incorporating a probability of event presence in this slot and its corresponding source location. To be consistent with the size of predictions, the ground truth in this application is padded to a set of size $N$ using a special ``no event'' label, which represents the absence of an event within a slot. For ``no event'' slot, its probability of presence is 0 and location coordinate is set to $(-1, -1)$ for a 2D case in our application. With these features, one of the main difficulties of training is to score predicted outputs with respect to the ground truth, particularly when dealing with multiple microseismic events with unknown numbers in the recorded segments.

A usual way to address the challenge of multiple events in one input segment is to consider the ordering of labels with respect to the individual events. \citet{wang2021direct} chose to concatenate the locations of multiple events in a predefined order (a.k.a., they ordered the events based on the arrival time) as a target sequence during training. However, the source locations of existing events are inherently an unordered set rather than an ordered sequence. Imposing a predefined order usually gives rise to the following intractable problems. First, it introduces bias in the training stage, which can highly penalize shifts in the order between locations \citep{ye2021one2set}. In other words, even if the network makes accurate predictions for each event, a shift in the ordering still results in a large training loss. Second, the predefined order increases the difficulty of training the network since the network needs to learn this special pattern that we have created.

In real-world scenarios, windowed segments of seismic recordings may contain zero, one, or several microseismic events. The actual number of events within one particular segment is often unknown, and perhaps more importantly, this number varies from one segment to another. To deal with this, the outputs of this application should be naturally expressed as sets of entities rather than vectors or sequences. In contrast to a vector or sequence, a set is a collection of elements which is invariant under permutation of the order of elements, and its size is not fixed in advance \citep{rezatofighi2017deepsetnet}. A pioneering solution for training a network which predicts a set is to find a bipartite matching between the ground truth and predictions, which is a typical linear sum assignment problem (LSAP). The Hungarian algorithm, a widely used algorithm in graph matching, is utilized for this assignment in our procedure. Given a cost matrix, the algorithm provides an optimal matching result in a way that the total matching cost is minimized \citep{li2022dn}. A loss function based on this remains invariant to changes in the ordering of predictions.

Let us denote $ y = \{y_i\}_{i=1}^N $ as the set of ground-truth events after padding, and $ \hat{y} = \{\hat{y}_i\}_{i=1}^N $ as the set of $N$ predictions. To find a bipartite matching between these two sets, we search for a permutation of $N$ elements $\sigma \in \mathfrak{S}_{N}$ with the lowest cost:
\begin{equation}
\sigma^* = \underset{\sigma \in \mathfrak{S}_{N}}{\arg \min} \sum_{i=1}^{N}\mathcal{L}_{\text{match}}\left(y_{i}, \hat{y}_{\sigma(i)}\right),
\label{eq:optimal_sigma}
\end{equation}
where $\mathcal{L}_{\text{match}}\left(y_{i}, \hat{y}_{\sigma(i)}\right)$ is a pair-wise matching cost between the ground truth $y_i$ and a prediction with index $\sigma(i)$. This optimal assignment can be completed efficiently through the Hungarian algorithm.

The matching cost takes both the classification accuracy and location error into consideration. The $i^{th}$ element in the ground truth set can be represented as $y_i = (c_i, s_i)$, where $c_i$ is 1 when an event is present in this slot and 0 for the ``no event'' case (i.e., a binary classification), and $s_i \in [0,1]^2$ is a vector that defines the normalized source coordinate location in reality, relative to the region of interest in a two-dimensional case. For the prediction with index $\sigma(i)$, we denote its probability of class $c_i$ as $\hat{p}_{\sigma(i)}(c_i)$ and its predicted source location as $\hat{s}_{\sigma(i)}$. With these notations, we define the matching cost as
\begin{equation}
\mathcal{L}_{\text{match}}\left(y_{i}, \hat{y}_{\sigma(i)}\right) = \norm{c_{i} - \hat{p}_{\sigma(i)}\left(c_{i}=1\right)} + \mathds{1}_{\left\{c_{i} \neq 0 \right\}} \mathcal{L}_{\text{MSE}}\left(s_{i}, \hat{s}_{\sigma(i)}\right),
\label{eq:matching_cost}
\end{equation}
where $\norm{\cdot}$ computes the absolute value of the element inside it, $\mathcal{L}_{\text{MSE}}(\cdot)$ is the mean squared error (MSE) loss function, $\mathds{1}_{\left\{c_{i} \neq 0 \right\}}$ represents an indication function which equals 1 when an event is actually present in the $i^{th}$ slot of the ground truth set and equals 0 otherwise.

\subsection{Network training: the Hungarian loss}

Based on the bipartite matching result, the so-called Hungarian loss is computed for all matched pairs to help the network optimize. The loss can be defined as a linear combination of the binary cross-entropy (BCE) loss for class predictions and the mean squared error loss for location predictions:
\begin{equation}
    \mathcal{L}_{\text{Hungarian}}(y, \hat{y}) = \sum_{i=1}^{N} \left[ \mathcal{L}_{\text{BCE}} \left(c_{i}, \hat{p}_{\sigma^*(i)}\left(c_{i}\right) \right) + \mathds{1}_{\left\{c_{i} \neq 0 \right\}} \mathcal{L}_{\text{MSE}}\left(s_{i}, \hat{s}_{\sigma^*(i)}\right) \right],
\label{eq:hungarian_loss}
\end{equation}
where $\sigma^*$ is the optimal assignment computed using Equation \ref{eq:optimal_sigma}. The use of the BCE loss is different from what we use in the matching cost, since the absolute difference between ground-truth and predicted probabilities offers a faster computation of the cost matrix.

\begin{figure}
  \centering
  \includegraphics[width=1.0\textwidth]{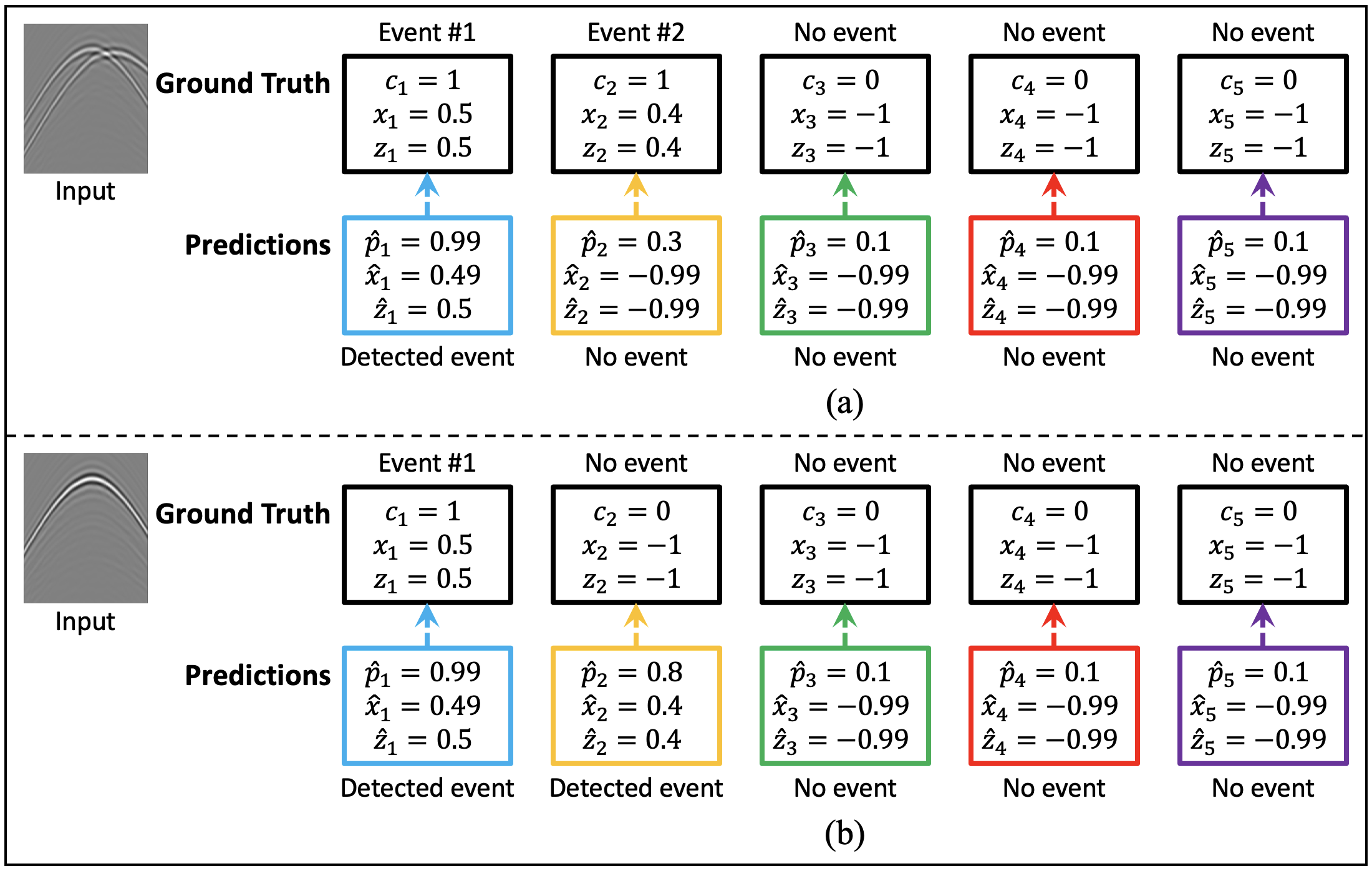}
  \caption{Illustration of two common scenarios where the predicted number of events does not match the ground truth. In case (a), the input segment contains two events, but the network only detects one event and misses the second. In case (b), the input segment contains a single event, but the network over-predicts to two events. The colored arrows indicate the optimal assignment determined by the Hungarian algorithm.}
  \label{fig:loss_penalty_cases}
\end{figure}

Fig. \ref{fig:loss_penalty_cases} illustrates two common scenarios where the predicted number of events does not match the ground truth, providing an intuition on the behavior of the loss function. The colored arrows indicate the optimal assignment determined by Equation \ref{eq:optimal_sigma}. In Fig. \ref{fig:loss_penalty_cases}(a), the input segment contains two events, but the network only detects one event while missing the other. The omission of the second event leads to a significant increase in the loss. The network is penalized through both the BCE loss term $\mathcal{L}_{\text{BCE}} (c_2, \hat{p}_2)$ from the classification part and the MSE loss term $\mathcal{L}_{\text{MSE}}((x_2, z_2), (\hat{x}_2, \hat{z}_2))$ from the location part. In Fig. \ref{fig:loss_penalty_cases}(b), the input segment includes a single event, but the network over-predicts to have two events. In this case, the non-existent event incurs a big penalty only from the BCE loss term $\mathcal{L}_{\text{BCE}} (c_2, \hat{p}_2)$ due to the wrong classification.

In practice, we notice that the classification loss term is computed much more frequently than the location loss term in the overall loss function. Given the difficulties presented to these two tasks, the classification loss term is down-weighted during implementation to compensate for the data and class imbalance. For instance, in our application, we found an empirical 1:9 ratio between the classification and location loss terms was required to balance their contributions effectively.

\section{Numerical Examples}

The proposed approach is first validated on synthetic passive seismic data as a conceptual proof, where we analyze its accuracy and efficiency in multiple-event detection and location. Afterwards, we apply this approach to a real-world microseismic dataset obtained from the field to attest its practical effectiveness.

\subsection{The SEAM Time Lapse model example}

\begin{figure}
  \centering
  \includegraphics[width=0.7\textwidth]{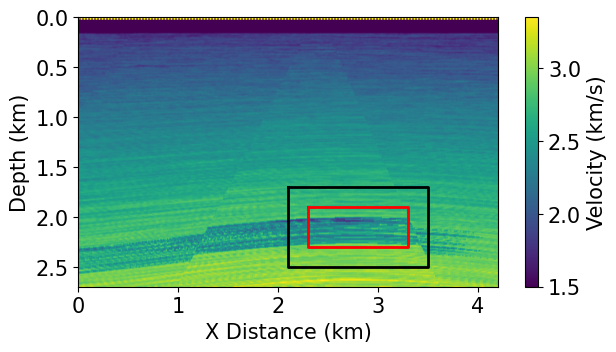}
  \caption{The SEAM Time Lapse model. The smaller red box represents the expected area of microseismicity and is used for generating test (assumed real) data, while the bigger black box is selected and used for generating training data. The receivers, denoted with yellow dots, are deployed on the surface at a regular interval of 30 meters.}
  \label{fig:seam_model}
\end{figure}

The SEAM Time Lapse model (Fig. \ref{fig:seam_model}) is chosen to imitate a scenario of $CO_2$ injection and microseismic event induction to test our proposed approach. The red box in Fig. \ref{fig:seam_model}, of size 1.0 by 0.4 kilometers, represents the expected area of potential microseismic activities. To mimic real conditions, as mentioned above, the training and validation data are generated using a smooth version of subsurface model, as if it was obtained from migration velocity analysis, while the test (assumed real) data are generated using the true subsurface model. Additionally, the training data are from sources within the black box in Fig. \ref{fig:seam_model}, which is a bigger area than the red box to allow for better predictions around the boundaries of target area, while the test data are from within the red box area. In this 2D example, a line of receivers are deployed on the surface as the monitoring array, which has 141 receivers (denoted with yellow dots) in total along x-direction with a regular interval of 30 meters. Based on this setup, 500 zero-event, 2000 one-event, 2000 two-event, and 2000 three-event seismic segment samples are generated to train the network. Detecting if there is any event or not is a relatively simple task for the network. Thus, substantially less zero-event training data are needed for the training. An additional benefit of this is that it helps to reduce the training cost. Each sample is made up of the preprocessed seismic data, labels of the number of events, and labels of their corresponding source coordinate locations. White Gaussian noise, as is conventionally applied in most benchmarking studies \citep{berkner1998wavelet, liu2017microseismic, wang2021direct}, is added to the training and validation data in between training iterations to mimic real-life scenarios where perfectly clean data are unavailable. The noise injection can also improve the network performance by allowing some inaccuracy when training networks \citep{luo2014deep}. During the inference stage, the same level of noise is also incorporated in the test data to achieve an SNR around 5. The SNR of a given window is calculated as the ratio of signal to noise energy, following a standard definition.

\begin{figure}
  \centering
  \includegraphics[width=0.9\textwidth]{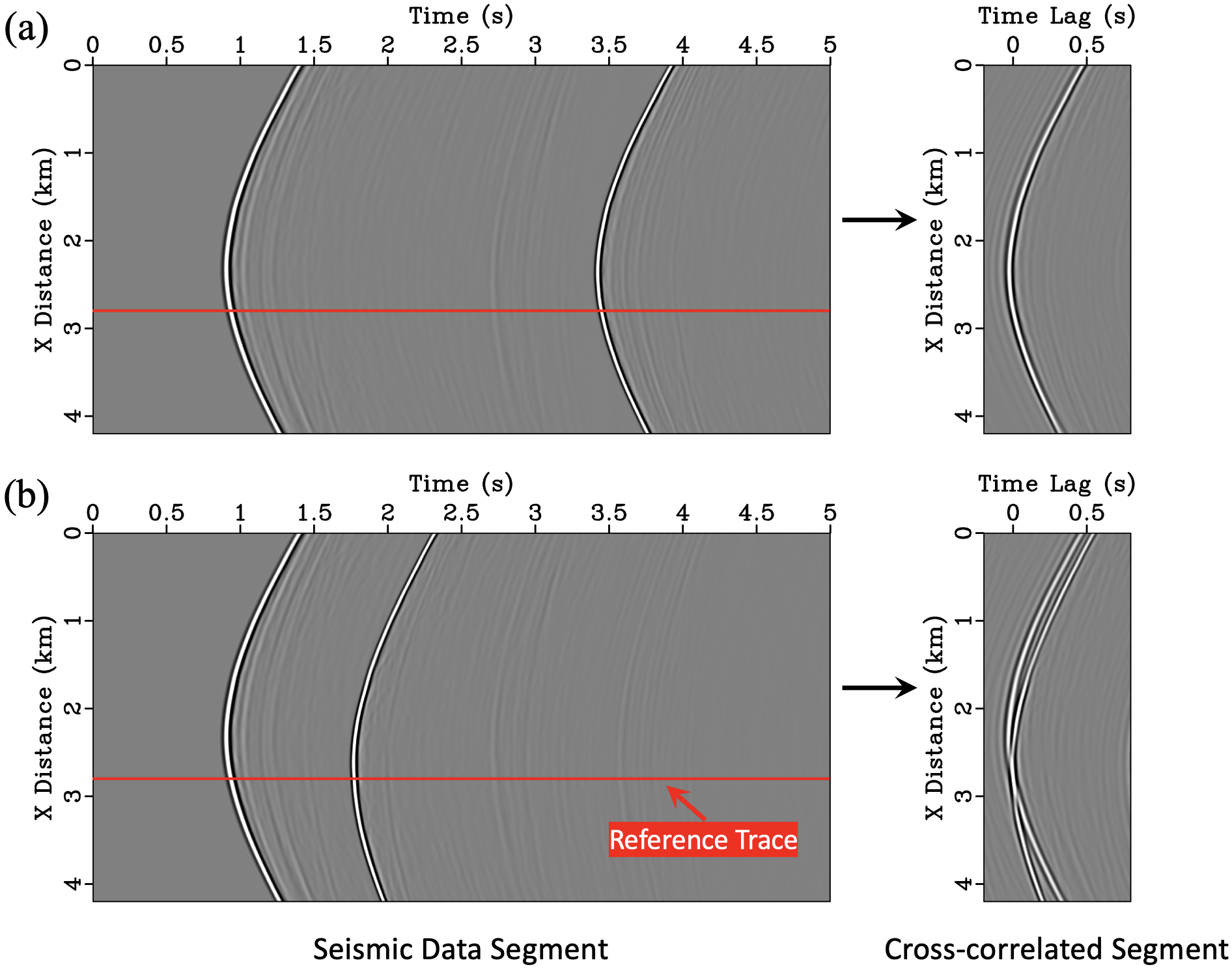}
  \caption{Passive seismic data segments before (left) and after (right) cross-correlation processing. In (a), two sources are positioned 50 meters away in both x and z directions, ignited 2.5 seconds apart. In (b), two sources are located 400 meters away in both directions, ignited 0.7 seconds apart. The red lines indicate the fixed-location reference trace used for cross-correlation.}
  \label{fig:source_proximity}
\end{figure}

The trained network demonstrates its competence in detecting microseismic events within the input seismic segments. Testing it on 1000 test samples from each class (i.e., zero-, one-, two-, and three-event cases, respectively), the classification accuracy is assessed via a confusion matrix as shown in Table \ref{tab:cls_confusion_matrix}, where the actual number of events in one input segment is compared against the predicted number. The values in the table represent the percentage of instances where each input segment was classified as having a particular number of predicted events. For example, in the row labeled ``1-event'': among the 1000 test samples where each contains a single microseismic event, 91.8\% of them were correctly classified as segments with one event in it, while 8.2\% of the input segments  were wrongly classified by the network as containing two events. In this layout, the diagonal elements represent the percentages of correct classifications, while the off-diagonal elements indicate the mis-classifications. Remarkably, the network achieves a flawless accuracy of 100\% for the simple task of event presence detection. Furthermore, the network consistently maintains over 90\% accuracy in classifying the number of events within an input segment across all cases. These results highlight the strong capability of this approach in accurately detecting microseismic events in the input seismic segment, under the specific conditions tested (including white Gaussian noise and an SNR around 5). Notably, the slightly lower classification accuracy observed in the one-event case is due to the fact that the two-event and three-event inputs may occasionally be interpreted as a single event by the network if the events' sources are closely located in space. For example, as shown in Fig. \ref{fig:source_proximity}(a), when two sources are located very closely in space, their event moveout patterns become very similar. If these two events fall within the same windowed time segment (even with a large difference in origin time), both of them will be moved to the area near zero-lag after the cross-correlation procedure (which is one of the required preprocessing steps) and thus are highly likely to overlap with each other. In this case, it is difficult for the network to distinguish between the two events. However, as shown in Fig. \ref{fig:source_proximity}(b), when two event sources are far apart in space, even with close origin times, their differing moveout patterns will reveal they are separate events (even after cross-correlation), as such the network is able to deal with multiple events within one input data segment. In other words, this network focuses primarily on the spatial distribution of events within each segment.

\begin{table}
\caption{Classification Results (Unit: \%)}
\label{tab:cls_confusion_matrix}
\centering
\begin{tabular}{|c|c|cccc|}
\hline
                              &                  & \multicolumn{4}{c|}{Predicted}                                                                                                          \\ \hline
                              &                  & \multicolumn{1}{c|}{\textbf{0-event}} & \multicolumn{1}{c|}{\textbf{1-event}} & \multicolumn{1}{c|}{\textbf{2-event}} & \multicolumn{1}{c|}{\textbf{3-event}} \\ \hline
\multirow{4}{*}{\rotatebox[origin=c]{90}{Actual}} & \textbf{0-event} & \multicolumn{1}{c|}{\textbf{100}}     & \multicolumn{1}{c|}{0}                & \multicolumn{1}{c|}{0}                & 0                \\ \cline{2-6} 
                              & \textbf{1-event} & \multicolumn{1}{c|}{0}                & \multicolumn{1}{c|}{\textbf{91.8}}    & \multicolumn{1}{c|}{8.2}              & 0                \\ \cline{2-6} 
                              & \textbf{2-event} & \multicolumn{1}{c|}{0}                & \multicolumn{1}{c|}{0}                & \multicolumn{1}{c|}{\textbf{93.5}}    & 6.5              \\ \cline{2-6} 
                              & \textbf{3-event} & \multicolumn{1}{c|}{0}                & \multicolumn{1}{c|}{0}                & \multicolumn{1}{c|}{1.6}              & \textbf{98.4}    \\ \hline
\end{tabular}
\end{table}

\begin{figure}
  \centering
  \includegraphics[width=1.0\textwidth]{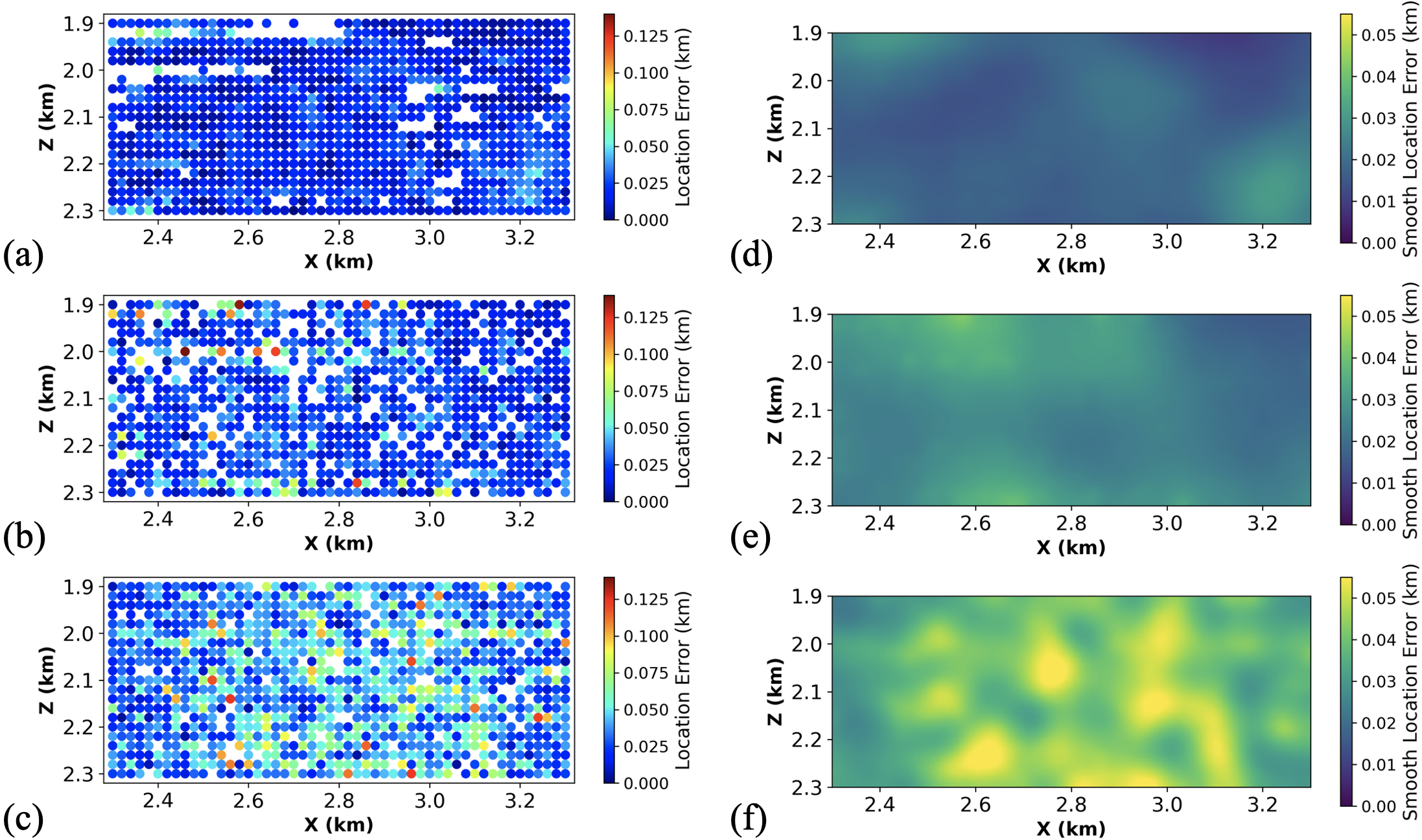}
  \caption{Source location maps of the test data with right classification predictions, colored with the actual location error values, for (a) one-, (b) two-, and (c) three-event cases, and the corresponding location error maps (after Kriging interpolation and Gaussian smoothing) for (d) one-, (e) two-, and (f) three-event cases. White spots in the left-sided maps indicate missed event detections at those positions.}
  \label{fig:seam_location_error}
\end{figure}

The accuracy of locating microseismic events back to their subsurface sources is assessed by computing the location errors, which quantify the Euclidean distances between the predicted and ground-truth source coordinate locations of the test data with correct classification predictions. Figs. \ref{fig:seam_location_error}(a)-(c) visually depict the maps of the sources located for one-, two-, and three-event cases, respectively, with the colormap reflecting the location error values. In the two-event and three-event cases shown in Figs. \ref{fig:seam_location_error}(b) and \ref{fig:seam_location_error}(c), two or three grid points are randomly selected as sources to generate multiple-event input segments. The location error at each grid point is the average prediction error from all input segments that include this point as one of their sources. The white spots in the maps indicate missed event detections at those positions. Figs. \ref{fig:seam_location_error}(d)-(f) present the corresponding location error maps after applying Kriging interpolation and Gaussian smoothing, to better illustrate the distribution of errors across the entire domain of interest. Additionally, the average location errors for one-, two-, and three-event cases are \textbf{18.4}, \textbf{25.6}, and \textbf{39.8} meters, respectively. As expected, the location errors increase as the number of events within the input segment grows, since the network is dealing with more complicated inputs where events may overlap or interfere with each other. This indicates that, for scenarios where multiple microseismic events are windowed in a relatively long time segment, the location accuracy is generally lower. It is worth noting that the area in which we placed the sources for generating the training data is of 1.4 by 0.8 kilometers. Therefore, the average location errors are all less than 5\% with respect to the size of the monitoring region. These results demonstrate the capability and reliability of this approach in accurately locating microseismic events under the current setup.

To provide a systematic comparison, we apply the least-squares diffraction stack migration method as a benchmark for microseismic event location, alongside the proposed approach. Like other conventional location techniques, this method is designed for single-event localization within the windowed time segments. Both this method and ours are tested on the same 1000 one-event time segments with an SNR of 5. The results, including the location error and average runtime, are summarized in Table \ref{tab:location_comparison}. With the known and exact origin time provided for each event in each segment, the least-squares migration method delivers highly accurate location estimates on average, though it occasionally exhibits significant outliers. In contrast, the proposed DETR network does not require knowledge of the origin time for localization, but also achieves a good location performance. Moreover, the considerable computational time of migration-based methods impedes their real-time applicability, while the DETR network offers a real-time solution. As shown, to process 1000 seismic segments, the least-squares migration method requires nearly two days of computation, whereas our network handles the same data in under 10 seconds, underscoring its efficiency for real-time applications. It's worth noting that the least-squares migration algorithm used in this study is not optimized at the industry level and could potentially be sped up significantly with a more efficient implementation. However, even with such optimizations, we believe it would still be unlikely for it to process 1000 segments within 10 seconds, as our method achieves.

\begin{table}
\caption{Comparison of location capabilities between the proposed DETR network and the least-squares diffraction stack migration (LSM) method for one-event seismic data segments.}
\label{tab:location_comparison}
\centering
\begin{tabular}{|c|ccc|c|}
\hline
                     & \multicolumn{3}{c|}{\textbf{Location Error (m)}}                                                 & \multirow{2}{*}{\textbf{Runtime (s)}} \\ \cline{1-4}
                     & \multicolumn{1}{c|}{\textbf{Average}} & \multicolumn{1}{c|}{\textbf{Minimum}} & \textbf{Maximum} &                                   \\ \hline
\textbf{DETR (ours)} & \multicolumn{1}{c|}{18.4}             & \multicolumn{1}{c|}{0.2}              & 77.6             & 0.007                             \\ \hline
\textbf{LSM}         & \multicolumn{1}{c|}{3.7}              & \multicolumn{1}{c|}{0.1}              & 131.2            & 150.0                             \\ \hline
\end{tabular}
\end{table}

\begin{figure}
  \centering
  \includegraphics[width=1.0\textwidth]{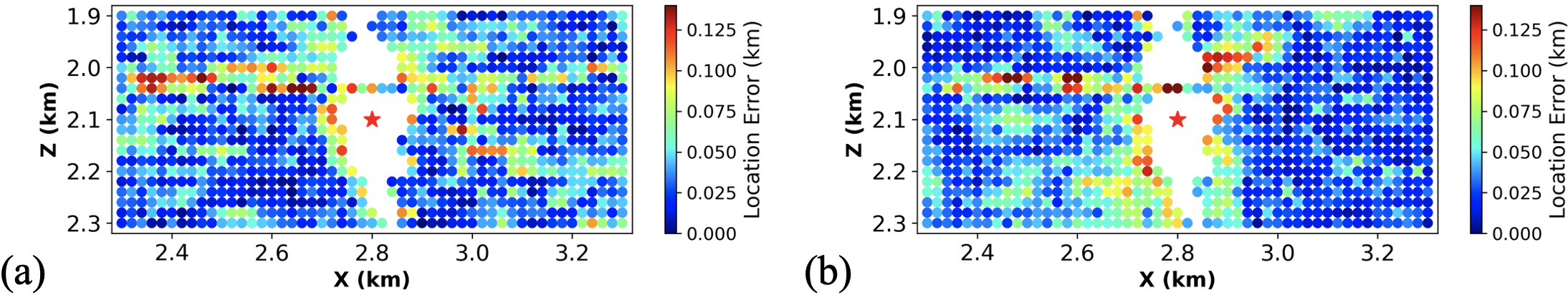}
  \caption{The source location error maps of 1070 two-event input segments, in which one source is placed at one of the 1070 regular gird points and the other source is placed at the center of this region (denoted with the red star). (a) displays the location errors for the various sources on the regular grid, and (b) displays the location errors of a single point (i.e., the central source) with being placed at the location of the grid point in which the other source is located (for better illustration). White spots indicate wrong event detections at those positions.}
  \label{fig:dist_sensitivity}
\end{figure}

To further explore the sensitivity of source location prediction to the distance between events within a single segment, we perform the following experiment. Over the whole region of potential source location defined earlier, a single source is placed on a regular grid with an interval of 20 meters across both axes to generate 1071 single-event data. Then, the single-event data for each source are added to the data from a source in the center of the domain to generate 1070 double-event data. By inputting these data to the network, location error maps are obtained for the double-event case. Fig. \ref{fig:dist_sensitivity}(a) displays the location errors for the various sources on the regular grid, while Fig. \ref{fig:dist_sensitivity}(b) shows the location errors of a single point, specifically the central source, placed at the location of the grid point in which the other source is located. The white region in the maps represents grid points with wrong one-event classification, indicating inputs where the network failed to distinguish the two events. Clearly, the distance between events significantly influences the location accuracy. The general trend observed is that closer proximity between events leads to lower location accuracy for both events, with the deeper event generally exhibiting better accuracy compared to the shallower one. Moreover, the sensitivity differs in the x- and z-axis directions. The location is better determined in x-direction. This makes sense since, based on the frequency range (5-20 Hz) and the offset (4.2 kilometers) under consideration, the data exhibit more significant differences when perturbing the event in the x-direction compared to the depth direction. This suggests that horizontal movement of events, which changes the location of its apex, is more significant than variations in curvature for the neural network. This feature, of course, depends on the frequency band and the aperture of recording.

\begin{figure}
  \centering
  \includegraphics[width=0.75\textwidth]{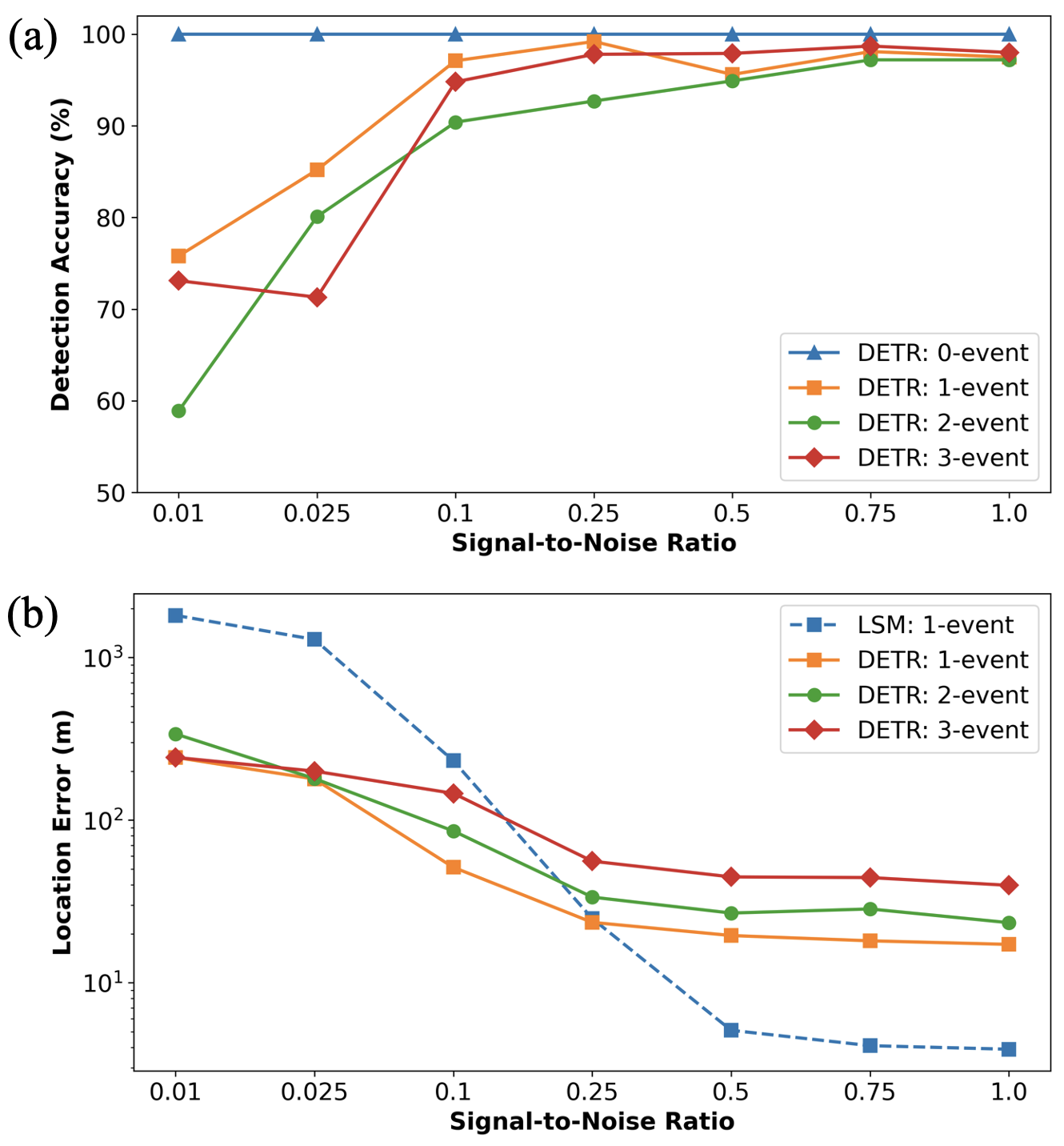}
  \caption{Results of the noise sensitivity study, evaluated by (a) detection accuracy for 0-, 1-, 2-, and 3-event input segments and (b) average location errors for 1-, 2-, and 3-event input segments from the proposed DETR network. (b) also includes the average location errors for 1-event input segments from the least-squares diffraction stack migration, for comparison.}
  \label{fig:noise_sensitivity}
\end{figure}

Given that noise is a critical component in microseismic systems, a preliminary noise sensitivity analysis is conducted to evaluate the network's robustness against varying noise levels. In this set of tests, white Gaussian noise is introduced to simulate different SNRs, ranging from 0.01 to 1 (i.e., 0.01, 0.025, 0.1, 0.25, 0.5, 0.75, and 1). As illustrated in Fig. \ref{fig:noise_sensitivity}(a), with the SNR decreasing, the detection accuracy declines for 1-, 2-, 3-event input segments. However, the consistent 100\% accuracy for 0-event input segments across all noise levels demonstrates that the network is highly capable of distinguishing seismic signals from white noise within the tested SNR range. Similarly, event localization accuracy, shown in Fig. \ref{fig:noise_sensitivity}(b), degrades as noise increases across all cases, with a noticeable failure at an SNR of 0.01, where the location errors become unacceptable. These findings indicate that the network's performance is significantly affected by noise contamination, particularly at low SNRs. The resilience of the proposed method to strong-noise environments still remains an important topic for future research and improvement.

For a more comprehensive comparison of noise sensitivity, Fig. \ref{fig:noise_sensitivity}(b) also includes the location results for 1-event input segments from the least-squares diffraction stack migration. In this case, we assume the accurate origin time is known for each event and used in migration. At higher SNRs, the least-squares migration method shows better performance in event localization compared to the proposed method, but definitely with the help of exact origin time information, which our method does not require. As the SNR decreases, its sensitivity to noise becomes apparent, and the method's reliability to locate events quickly decays, making it hard to rely on in strong-noise environments. In contrast, our proposed method exhibits better resilience under low SNR conditions, highlighting its robustness and potential in real-world applications with massive noise interference.

\subsection{The Arkoma Basin field data example}

The proposed approach is further validated on a passive seismic dataset recorded for the purpose of hydraulic fracture monitoring of a shale gas reservoir in the Arkoma Basin, North America. The Arkoma Basin, an east-west elongated sedimentary basin stretching through Oklahoma and Arkansas, is one of the most well-known gas production areas in America. At this site, a total of 911 single-component vertical geophones were deployed on the surface for microseismic monitoring, which spread out along 10 radial lines centered at the wellhead. More information about this dataset can be found in previous work on the same dataset by \citet{stanvek2017seismicity}.

As a conceptual proof of the proposed approach, instead of using a 3D subsurface velocity model, we have chosen to initially work with a relatively simple 2D case and three previously identified field events on this 2D plane to investigate the applicability of this approach on the field data. The extension of this approach to 3D cases is further discussed in the Discussion section, where we address potential challenges and solutions for applying the method in real-world 3D scenarios. For our current 2D network, only one arm of receivers (122 receivers) is taken from the star-like arrays as well as the corresponding 2D velocity model under this receiver line. With an average receiver spacing of 23 meters, this selected 2D line spans around 2800 meters of length.

\begin{figure}
  \centering
  \includegraphics[width=0.9\textwidth]{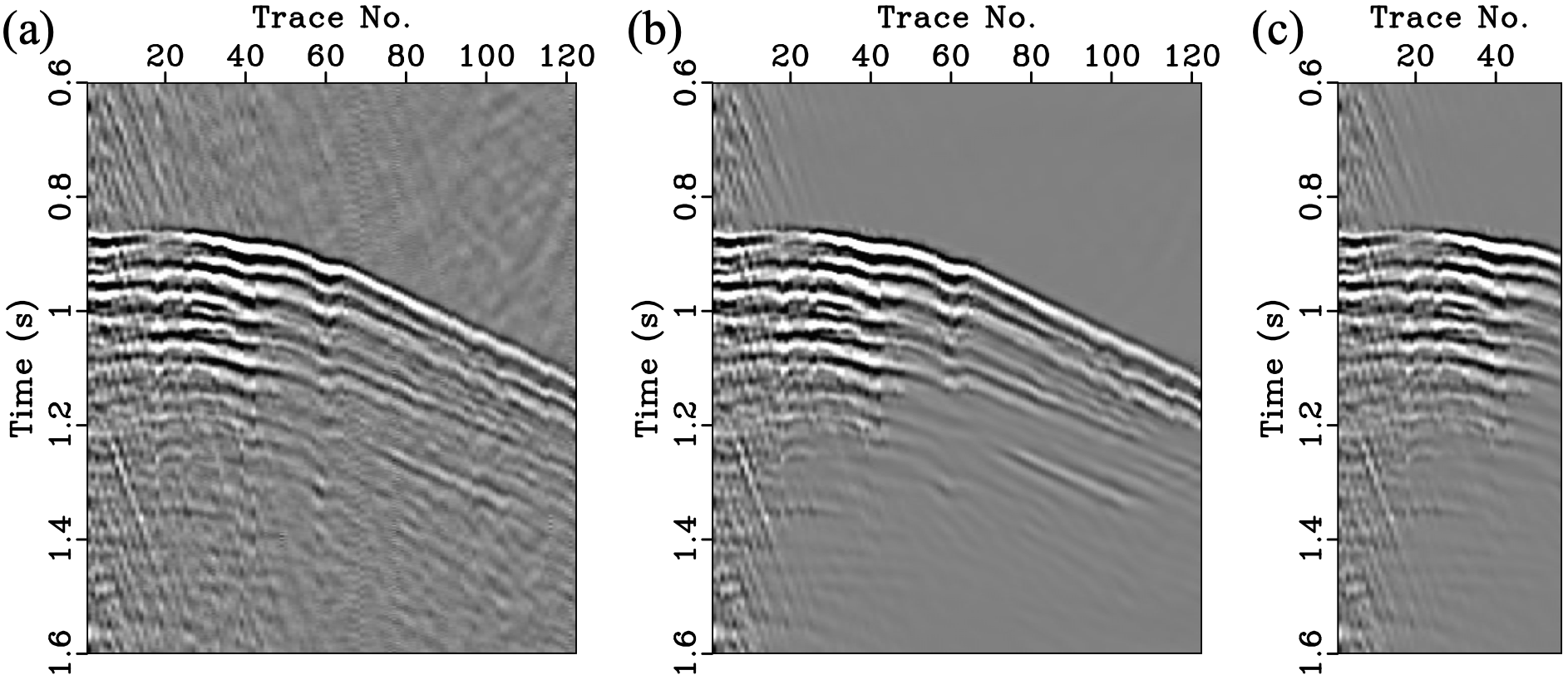}
  \caption{Examples of (a) original field data, (b) denoised field data, and (c) processed field data after removing the traces corresponding to the linear moveout.}
  \label{fig:arkoma_field_data}
\end{figure}

An exploratory data analysis is first conducted on the field data, an example of which is shown in Fig. \ref{fig:arkoma_field_data}(a). To improve the data quality to enhance the effectiveness of the subsequent pipeline, a simple non-local means method is employed for denoising, as illustrated in Fig. \ref{fig:arkoma_field_data}(b). Across all the field events, it is observed that the larger-offset data, which is the right side of the recordings shown in Fig. \ref{fig:arkoma_field_data}(b), exhibit a moveout pattern substantially more linear than the usual hyperbolic shape, and we propose these events have passed through a high velocity medium not present in the provided velocity model \citep{anikiev2014joint}. The calculated propagation velocities for this linear moveout consistently exceed 6500 meters/second. However, the maximum velocity of the subsurface model derived from an active seismic acquisition and calibrated with four perforation shots \citep{anikiev2014joint} is only 5350 meters/second. This indicates that the linear moveout can be attributed to other subsurface objects or layers, which are not incorporated in the provided subsurface velocity model.

By training with the synthetic passive seismic data, the velocity information of the subsurface model is embedded in the network for instant event location. More specifically, the temporal-spacial distribution of microseismic events is determined by the medium velocities that the seismic waves pass through, which is also in turn utilized to estimate the event locations by the network. According to the general trend of velocity increasing with depth inside the earth due to the nature of sedimentation, deeper areas in the subsurface often correspond to larger velocities, which result in smaller curvatures for events from these area. Therefore, when inputting the linear moveout part of the data to the network, which is trained to perceive that all data travel directly from the source, the network will reconcile the small curvature (linear moveout) as if it originated from a significantly deeper source. Considering that the proposed method is sensitive to the event curvature as indicated in the synthetic sensitivity analysis, the traces corresponding to the linear moveout are removed. The resulting processed field data are presented in Fig. \ref{fig:arkoma_field_data}(c).

\begin{figure}
  \centering
  \includegraphics[width=1.0\textwidth]{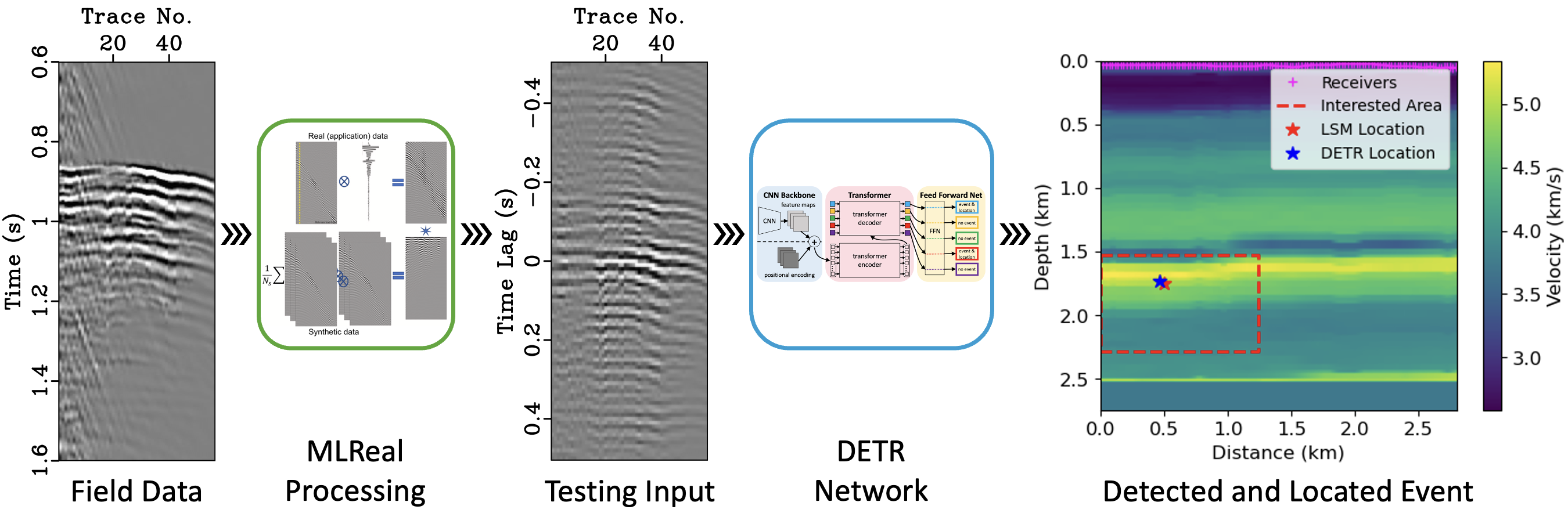}
  \caption{The inference workflow for one of the field test data, where one event is present in it.}
  \label{fig:arkoma_inference_workflow}
\end{figure}

Since only windowed data segments with single field events from the Arkoma Basin are provided, an experiment is carried out for the scenario of single-event detection and location to evaluate the efficacy of the proposed approach on field application. As such, the training data contain either zero or one event. The network is trained using 2000 synthetic one-event passive seismic segments that are generated based on the provided subsurface model, obtained from a previous active seismic survey \citep{stanvek2017seismicity}. These synthetically generated microseismic signals are generally hyperbolic across all the 122 traces (without linear moveout). Therefore, to be consistent with the field data that are considered reliable for event location, all the synthetic training data consist of 56 traces spanning the same range as field data. The training set also includes an additional 500 zero-event seismic time segments. For testing purposes, three field seismic data segments are selected with previously detected and analyzed events \citep{anikiev2014joint, stanvek2015semblance, stanvek2017seismicity} (each segment with one event within it), and their reference locations were earlier determined using the provided velocity model and all 911 geophones in 3D. By comparing the predicted source locations from the network with the reference locations of these field events, we obtain location errors of 48.7, 26.0, and 60.2 meters, respectively. The inference workflow for one of the field test data is illustrated in Fig. \ref{fig:arkoma_inference_workflow}. Given that the whole area for potential microseismicity is as large as 1250 by 750 m and only a short array with poor illumination is utilized for event location, these results showcase the impressive performance and accuracy of this approach when applied to real field data for detecting and locating microseismic events from the subsurface. More adequate data acquisition and better illumination, like longer seismic arrays, are likely to improve the location accuracies from the seismic recordings, assuming that they are accompanied by a suitable velocity model. In this case, the execution time of the network for predicting on each segment is approximately 6.9 milliseconds. Therefore, with better-suited data, this method could have been applied for real-time microseismic event detection and location, achieving results within milliseconds per time segment.

Additionally, the commonly used least-squares diffraction stack migration method is applied here to provide a useful benchmark to compare the proposed method against. The benchmark locations, which were determined from migration using 56 traces, deviate by 42.6, 70.7, and 15.8 meters from the predicted locations of the trained network. Concurrently, these benchmark locations have discrepancies of 91.2, 78.9, and 45.4 meters with respect to the reference locations. Notably, while inputting the full 122 traces for location, it's consistently observed that the depth estimations from diffraction stack imaging are always much deeper (generally 200-300 meters) than those in reference locations. This further validates the necessity of excluding the higher-offset traces from the field data when inputted for location determination to enhance the estimation accuracy.

\begin{figure}
  \centering
  \includegraphics[width=0.75\textwidth]{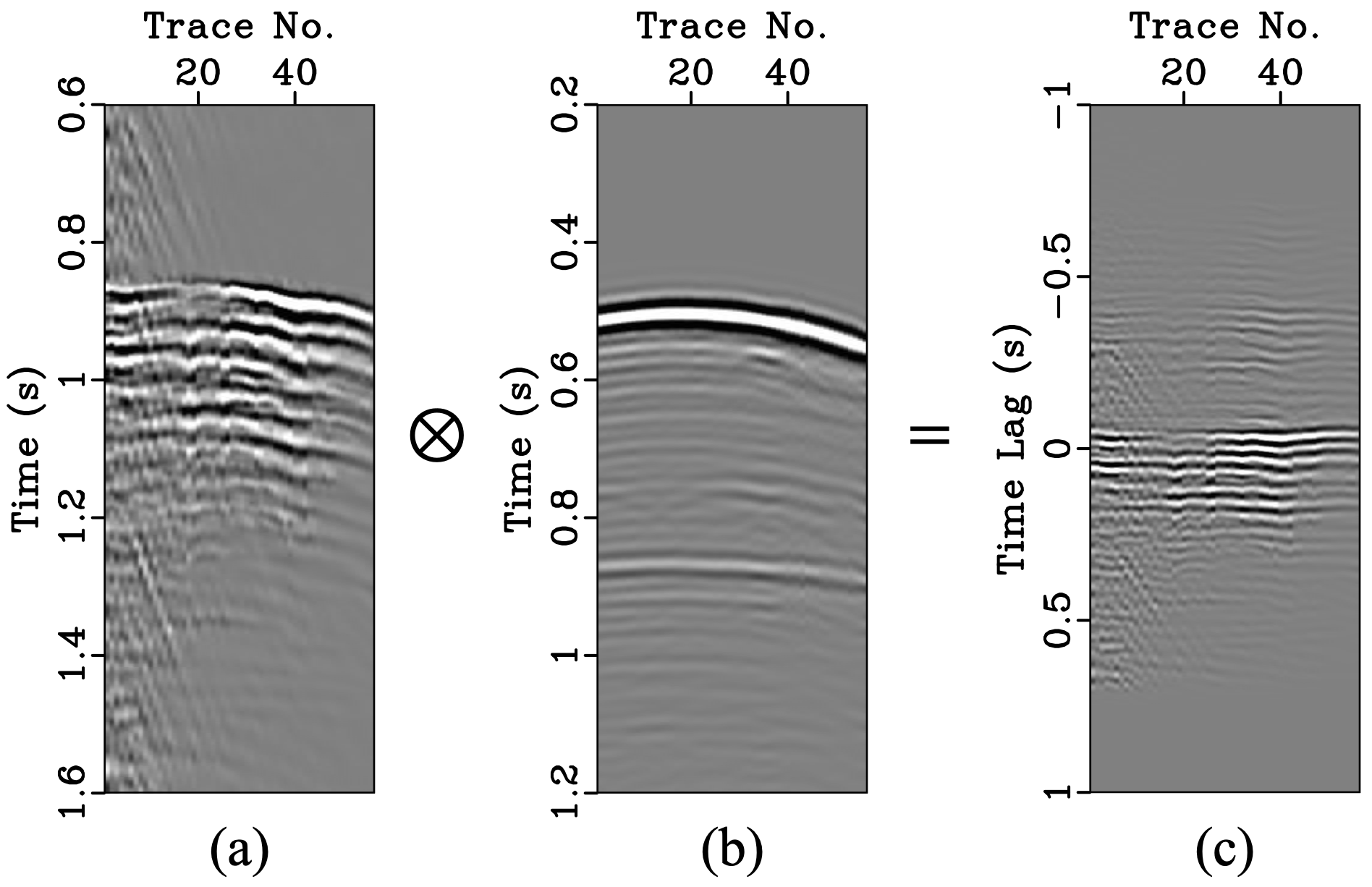}
  \caption{Cross-correlation (symbolized by $\otimes$) between (a) the field data and (b) the synthetic data generated from the corresponding predicted source location, resulting in (c) the cross-correlated data. The flatness shown in the cross-correlated data indicates a high degree of agreement in waveform moveout between the field and synthetic data, which further proves the accuracy of the predicted source location.}
  \label{fig:arkoma_xcorr}
\end{figure}

As an in-depth validation of the accuracy of the network's prediction, synthetic data are generated from the predicted source location (Fig. \ref{fig:arkoma_xcorr}(b)). The synthetic data exhibit nearly identical moveout to the observed field data shown in Fig. \ref{fig:arkoma_xcorr}(a). To quantify this similarity, a cross-correlation is performed between them, with the resulting cross-correlated data displayed in Fig. \ref{fig:arkoma_xcorr}(c). The observed flatness in the cross-correlated data confirms a strong agreement in waveform moveout between the field and synthetic data, which directly supports the accuracy of the location estimate provided by the network.

\section{Discussion}

\subsection{Utilization of MLReal}

In this study, we proposed to train the network on synthetic seismic data and apply it to field seismic data. Therefore, the level of similarity between the training and application datasets matters. Often, the synthetic data fail to capture features of the field experiment, like a proper waveform source signature and the ambient noise. Thus, we easily end up with poor performance of trained ML models at the inference stage. This problem in our application is solved through the MLReal domain adaptation technique, which has also been successfully applied in other work \citep[e.g.,][]{birnie2022leveraging, zhang2022improving}. By transforming the distributions of training and application datasets into much closer ones through a series of cross-/auto-correlations and convolutions, MLReal helps the trained models generalize better on field data.

While applying MLReal, the time-axis information is eliminated by the correlation operation. Bringing all the events inside the input segment to the area near zero lag makes the multi-event detection and location harder for cases where events come from very close source locations and therefore exhibit similar moveout patterns. Usually, this might not be an issue for microseismic clustering applications where the shape and extent of the microseismic event cloud is more important than the exact number of events. However, if the number of events is important, one way to deal with this situation is to slice the original recording into shorter time segments where the likelihood of each segment containing at most one event is high.

\subsection{Real-time monitoring}

Although the examples shown above are both based on the surface passive seismic recordings, the proposed approach is designed to be generalizable and feasible in both surface and borehole monitoring systems, or even a hybrid system. This is because the network builds the projection from the patterns of microseismic energy across sensors to responsible source locations, based on the governing physics which is implicitly embedded into the training data and corresponding labels.

With our proposed method, the heavy computational burden is charged during the network training stage, which is offline. At the inference stage, all the processing procedures on the field data (including segmentation, cross-correlation, and resampling) are very efficient to apply. More importantly, the inference time of each input data is on the order of $10^{-3}$ seconds. Thus, once the network is adequately trained, it can be applied for joint microseismic event detection and location with a real-time completion. In addition to this, it can be in service for a long time (e.g., for months) at the cost of one-time training. Moreover, once the subsurface velocity model or acquisition geometry is updated, the network can also be efficiently updated or calibrated using the technique of transfer learning for better estimations. This ensures that engineers and scientists are informed of the locations of events in the field instantly. For incorporating MLReal, we have to utilize some field data in the training, which can be done over a short period of time. Once enough features of the real data are incorporated in the training, the application of the network for real-time monitoring should be possible.

Previously, \citet{wang2021direct} proposed the use of two separate networks for simultaneously detecting and locating microseismic events. In such cascading systems, the input data and information go from one network to another. However, tasks like event detection and location make use of the same features in the input seismic data, which is the waveform. \citet{mousavi2020earthquake} illustrated that for the scenario of earthquake monitoring, combining these two tightly-related tasks jointly in a single framework improves the model performance in each individual task. Moreover, feeding (almost) the same input data to two different networks and train them separately is generally inefficient. Finally, the application of an end-to-end framework will be faster than two networks in tandem for inference (assuming similar complexity of the network architectures).

In real-world applications, there are always scenarios where the field data have complications out of researchers or engineers' understanding and exhibit features that don't fit models of the local subsurface/geomechanics, like the linear moveout occurring in the Arkoma Basin field data. For the aforementioned linear moveout, as post-event analysis, the incorporation of a denoising procedure is recommended to remove unwanted waveforms. Whilst for real-time applications, we would recommend adapting the training data to incorporate such unwanted waveforms and teaching the network to ignore their presence. The incorporation of linear noise into training data was recently proposed by Chammaro et al. 2024 for surface wave analysis. The benefit of including such waveforms for microseismic training datasets is a topic for future study.

\subsection{Future work}

In this study, we have proven the value of the DETR framework assuming a single receiver line. However, surface microseismic monitoring arrays are typically two-dimensional, resulting in 3D datasets of $x,z,t$. Incorporating the additional spatial axis is likely to further improve our procedure due to the additional signal information available. For 3D applications, our approach doesn't demand densely and uniformly sampled seismic data. For the extension to 3D, two choices are available: (1) concatenating all the sensor lines in a horizontal distance axis and taking it as a 2D image as the input to the network, which requires no modification to the current network architecture; or (2) taking different 2D lines as different channels to the CNN backbone, as performed by \citet{wang2021data}, which is also implementation-wise simple since it only requires a change in the number of input channels in the CNN backbone from one to the number of sensor lines in the survey.

Another extension would be the transformation from an acoustic assumption to an elastic one. This could be easily achieved by changing the velocity model used for generating the synthetic training data. Next, by substituting class loss term from the binary cross-entropy loss to the multi-class cross-entropy loss function, this approach will naturally learn to handle P- and S-wave events within the seismic recordings. Apart from being closer to the real-world situations, more signals with their corresponding features in the seismic segments would hopefully improve the accuracy of the trained network.

At the current stage of this work, a known velocity model is expected for synthetic data generation, such that the network can be trained for accurate and reliable location of microseismic events. Poor velocity models will result in poor location estimations. Therefore, the common challenge of a poorly calibrated velocity model will be considered as a part of the future research.

The above extensions will enhance the importance of polarity flipping of seismic data. In reality, the polarity flip related to source radiation patterns is an inherent feature in passive seismic data. Therefore, to deal with real-life passive seismic data, the consideration of polarity flip is necessary. However, the polarity is often not required for source location of microseismic events. Some solutions may involve working on absolute amplitudes/energy, envelopes, similarity, or semblance of signals, like a number of conventional methods \citep{gajewski2007localization, eisner2008noise, grandi2009microseismic, chambers2010testing, gharti2010automated}. Such an approach can be included as a simple pre-processing step before passing the input data to networks for both training and inference to overcome source mechanism effects in future field applications of the proposed method.

\section{Conclusions}

We proposed an approach adapted from the DEtection TRansformer for joint microseismic event detection and location directly applied on the recorded seismic waveforms. The proposed network is built on a CNN backbone with a Transformer encoder-decoder block and trained using a set-based Hungarian loss. A synthetic study validated the strong capability of the proposed procedure. Furthermore, the procedure was applied to a field dataset from the Arkoma Basin, which evaluated its practicability and accuracy, and identified the acquisition conditions under which we can obtain optimal performance. Once trained, the network offers an end-to-end solution for real-time monitoring to detect and locate microseismic events, assuming a reliable velocity model and relatively good-quality data. Due to its ability to analyse varying number of microseismic events on-the-fly, we believe this work will pave the way for a new generation of real-time ML algorithms for joint microseismic event detection and characterisation.

\section*{Acknowledgments}
This publication is based on work supported by King Abdullah University of Science and Technology (KAUST). The authors thank the DeepWave sponsors for their support and Microseismic Inc. for the use of the Arkoma Basin field data. Appreciation also goes to Dr. Xinquan Huang and Dr. Hanchen Wang for the useful discussions and suggestions.

\bibliographystyle{gji}
\bibliography{references}

\end{document}